\documentclass[preprint,authoryear,12pt]{elsarticle}
\usepackage{graphics}
\usepackage{amssymb}
\usepackage[nodots]{numcompress}
\usepackage{lineno}

\journal{Icarus, accepted}

\begin{document}

\begin{frontmatter}

\title{Survey of Kozai Dynamics Beyond Neptune}

\author{Tabar\'{e} Gallardo\corref{cor1}}
\ead{gallardo@fisica.edu.uy}

\cortext[cor1]{Corresponding author}

\author{Gast\'{o}n Hugo}
\author{Pablo Pais}

\address{Departamento de Astronom\'{i}a, Instituto de F\'{i}sica, Facultad
de Ciencias, Igu\'{a} 4225, 11400 Montevideo, Uruguay}

\begin{abstract}

We study the Kozai dynamics affecting the orbital evolution of trans-neptunian objects being
captured or not in MMR with Neptune. We provide energy level maps of the type $(\omega,q)$ describing the possible orbital paths from Neptune up to semimajor axis of hundreds
of AU.
The dynamics for non resonant TNOs with perihelion distances, $q$, outside Neptune's orbit, $a_N$, is quite different from the dynamics of TNOs with $q<a_N$, already studied in
previous works.
While for the last case there are stable equilibrium points at $\omega =0^{\circ}, 90^{\circ}, 180^{\circ}$ and $270^{\circ}$ in a wide range of orbital inclinations, for the former case it appears a family of
stable equilibrium points only at  a specific value of the orbital inclination, $i \sim 62^{\circ}$, that we call critical inclination.
We show this family of equilibrium points is generated
by a mechanism analogue to which drives the dynamics of an artificial satellite perturbed by an oblate planet.
The planetary system also generates an oscillation in the longitude of the perihelion of the TNOs  with
$i \sim 46^{\circ}$, being Eris a paradigmatic case.
We discuss how the resonant condition with Neptune modify the energy level curves and the location of equilibrium points.
The asymmetric librations of
resonances of the type 1:N generate a distortion in the energy level curves and in the resulting  location of the equilibrium points in the phase space $(\omega,q)$.
We study the effect on the Kozai dynamics due to the diffusion process in $a$ that occurs in the Scattered Disk. We show that a minimum orbital
inclination is required to allow substantial variations in perihelion distances once the object is captured in MMR and that
minimum inclination is greater for greater  semimajor axis.

\end{abstract}


\end{frontmatter}


\section{Introduction}
\label{intro}

One of the open problems of the trans-neptunian region (TNR) is to
find an explanation for
the wide variety of orbits
that the discovered trans-neptunian objects (TNOs) exhibit. In particular, those objects
with perihelion outside Neptune's orbit and with
high orbital inclinations and eccentricities  that
 cannot be explained by the diffusive process the Scattered Disk Objects (SDOs) experience, nor by the
 past dynamical history of the planetary system
 \citep{nice1}.
It is known \citep{du95,gl02,fgb04} that TNOs with perihelion $q<36$ AU are inside a chaotic region that generates a diffusion in semimajor axis enhancing aphelion
 distances. But
the diffusion itself cannot decouple the perihelion from Neptune's region, then it is necessary to invoke another mechanism to explain the
existence of high perihelion SDOs, also known as Detached Objects.
The  origin of these problematic orbits have given rise to theories involving
passing stars \citep{id00},
scattered planets \citep{glch06},
tides from the star cluster where the Sun was formed \citep{brass06},
a stellar companion \citep{goal06}
and others.
But, in order to avoid extra hypothesis
it is necessary
an abroad panorama of the dynamics generated by the planetary system itself in the TNR.
Secular resonances, mean motion resonances (MMR) and the Kozai resonance (KR) are dynamical mechanisms  operating in the TNR and that could explain some eccentric  orbits decoupled from encounters with Neptune
\citep{fgb04,go05,go11}.
The present work contributes with new results following that line of thinking.

Secular resonances do not affect the TNOs located beyond $a\sim 42$ \citep{kz91} and MMRs with Neptune do not generate themselves
a substantial variation in the orbital elements, as for example the MMRs with Jupiter do \citep{ga07}.
The orbital variations observed when a particle is captured in MMR with Neptune are due to a secular dynamics inside the MMR, like Kozai dynamics.
The Kozai dynamics, or more properly Kozai-Lidov dynamics, has its roots in a study by  \citet{li62} of the secular evolution
of an artificial Earth's satellite perturbed by the Moon.
 Applying these ideas to the asteroid belt, \citet{ko62} developed an analytical approximation for the mean secular perturbing function due to Jupiter assumed in circular orbit, valid even
for high inclination and high eccentricity asteroidal orbits.
The conservation of the semimajor axis, as is usual in secular theories, but also the conservation of the parameter $H=\sqrt{1-e^2}\cos i$,  and the existence of an energy integral, $K(\omega,e)$, allowed the calculation
of the energy level curves for this model showing large oscillations of $e$ and $i$ coupled with the argument of the perihelion, $\omega$, around equilibrium points that appear for $H<0.6$ at $\omega = 90^{\circ}$ and $270^{\circ}$.
These maps of energy level curves have axial symmetry with respect to
 $\omega = \mathrm{k}\, 90^{\circ}$ being $\mathrm{k} = 0, 1, 2, 3, 4$ and
 with respect to
 $i = 90^{\circ}$. The orbital regime under oscillations of  $\omega$ it is known since then as Kozai resonance (KR) and it provides
  an explanation to the existence of some asteroids in spite of their large orbital oscillations.

\begin{figure}[]
\resizebox{12cm}{!}{\includegraphics{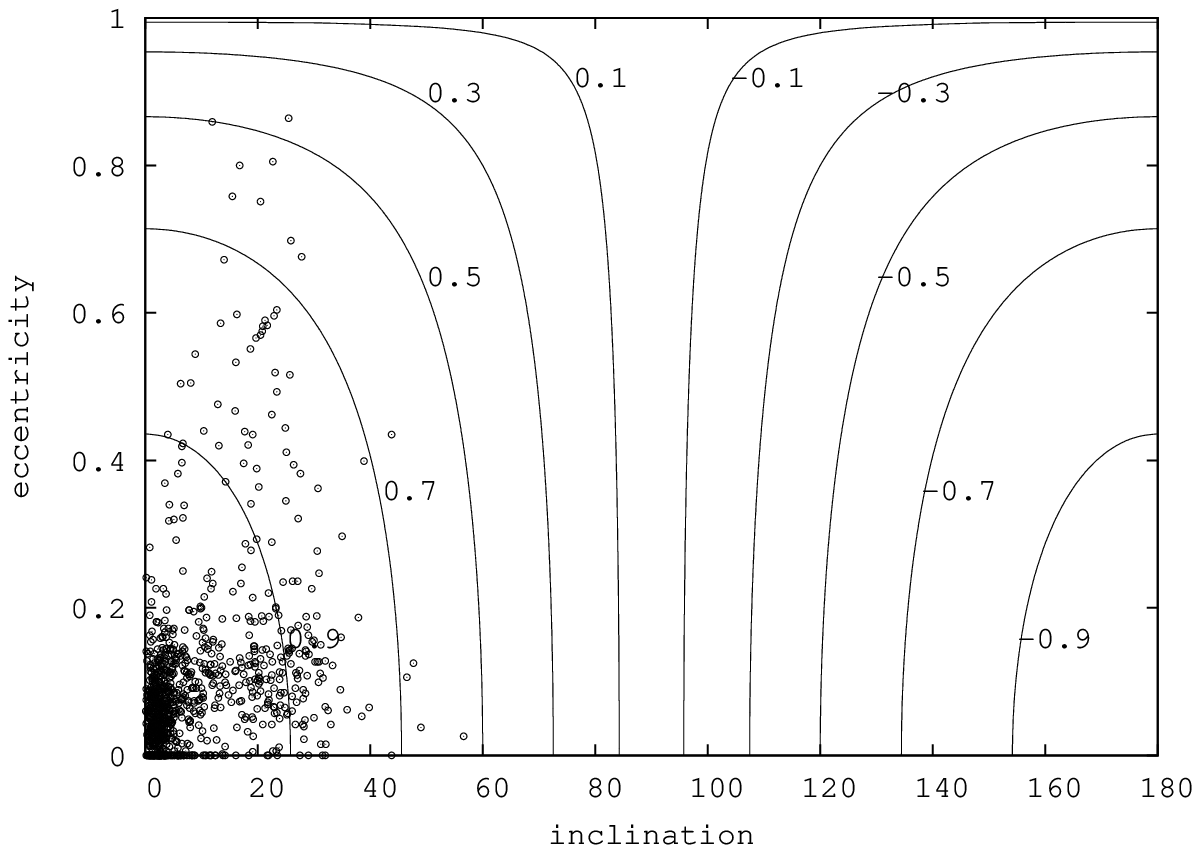}}
\caption{Trajectories of constant $H$ (labeled lines) and the known population of  objects with $q>36$ AU (open circles).
 All these objects could  increase or decrease their $q$ and $i$ by Kozai dynamics
 following curves of constant $H$.  Data corresponding to 986 objects from JPL, by January 2012.}
\label{real}
\end{figure}

The conservation of $H$ gives us the ultimate limits in $e$ and $i$ between which the particle's orbit can evolve assuming a secular dynamical regime, that means no close encounters with the planets, whether resonant or not. But the actual limits in $e$ and $i$ for a specific orbit are imposed by the orbital variations allowed by the energy level curve
to which the particle is confined.
We show in Fig. \ref{real} the known population of TNOs with $q>36$ AU, that means outside the diffusion region, in a diagram $H(i,e)$.
These objects at present do not undergo encounters with Neptune and do not experience diffusion in semimajor axis, then their eccentric or very inclined orbits are due to the early history of the planetary system or to the secular dynamics acting at present.
Each object could evolve along its corresponding $H$-constant curve allowing to increase or decrease its present perihelion distance, but the limits only can be defined  when studying the energy level curve for each object.

\citet{ko85}  modified his model to include also a mean motion resonant condition between the mean longitudes of Jupiter and the asteroid assuming a fixed value
of the critical angle that characterizes the resonance, allowing to understand the KR inside a MMR (or MMR+KR). Energy level curves for resonant asteroids are very different than for the non resonant case, and also are different the limits for the orbital oscillations. Moreover, for some resonances, new stable equilibrium solutions appear for  $\omega$ different from the known $90^{\circ}$ or $270^{\circ}$. These were called \emph{asymmetric} equilibrium points some years after and they appear as stability islands between collision trajectories with the planets \citep{gm99}. Except for the case of the Jupiter's Trojans, the energy level diagrams exhibit, as in the non resonant case, symmetry with respect to $\omega = \mathrm{k}\, 90^{\circ}$.

\citet{ba92} studied the orbital evolution of high inclination long period comets with a new semi-analytical averaging method that takes into account
 the perturbations of Jupiter assumed in circular orbit. The method is not based on any analytical series expansions but in
  calculating numerically the mean energy integral, doing it good for any set of orbital elements. They numerically computed energy level curves which long period comets should
 follow assuming no planetary close encounters or MMRs with Jupiter. That energy level curves showed stable equilibrium
 oscillations around $\omega = \mathrm{k}\, 90^{\circ}$ for cometary perihelion inside the planetary region. The principal result was
 that some comets have extreme perihelion oscillations which is clearly related to the origin of some sungrazing comets.

\citet{tm96} followed an analogue method but including the four jovian planets in order to study the dynamics in the outer Solar System. The series of energy level curves they obtained
give a very complete panorama of the perihelion behavior from the Neptune region down to the proximity of the Sun. They also obtained,
as did \citet{ba92},
the same equilibrium oscillations  around $\omega = \mathrm{k}\, 90^{\circ}$ for the case of long period comets. Nevertheless, by their detailed study, it is clear now that
the equilibrium islands at  $\omega = 90^{\circ}, 270^{\circ}$ with very small $q$ are more stable than the ones at $\omega = 0^{\circ}, 180^{\circ}$ which vanish in some circumstances due to the proximity with the collision curve with the planets, which is a curve  in $(\omega,e)$ that implies orbital intersection between the particle and the planets for a given value of $(a,H)$.
\citet{mt96} followed the same method of \citet{tm96} but including the effects of all the planets allowing the study of the dynamics of asteroids with $a<2$ AU in NEA type orbit, showing by first time the Kozai dynamics associated with the terrestrial planets.
\citet{kn99} and \citet{kn07} finally provided analytical solutions for the original Kozai model of an asteroid under an external perturber.

\citet{fgb04} and \citet{go05} in a series of numerical simulations showed the relevance that the interaction between MMRs and KR have for increasing significantly the perihelion
distances of SDOs, from $q\sim 40$ AU to $q\sim 70$ AU. They found the KR generates large perihelion variations only when the TNO is captured in MMR with Neptune
and preferably in resonances of the type 1:N.
\citet{go05} also provide figures showing how the energy level curves are modified when a MMR condition is imposed.

\citet{ga06} using a classical expansion up to order 10 of the disturbing function due to Neptune analyzed how the KR is generated inside and outside a MMR with Neptune. It was found a new regime of the KR for orbits with perihelion outside Neptune in the far TNR: it appears  a stable equilibrium point at $\omega = 90^{\circ}$ and $270^{\circ}$
but for a very particular value of the inclination, $i \sim 63^{\circ}$, called critical inclination, which resembles the dynamical behavior of Molniya like artificial satellites.
Curiously, \citet{ku02} had also detected notable eccentricity variations in fictitious  particles with initial circular orbits with $i=61^{\circ}$ at $a=43$ and $a=45$ AU but
with $\omega$ oscillating around $0^{\circ}$ and $180^{\circ}$.
Then, it is evident there is an interesting  Kozai dynamics related to orbits with $i \sim 61^{\circ} - 63^{\circ}$ in the TNR.
\citet{wh07} using the same classical expansion used in \citet{ga06} due to Neptune studied the effects of the KR inside the MMRs 2:3 and
1:2 providing energy level diagrams for that resonances, but they did not consider inclinations greater than $55^{\circ}$.
We must take into account that \citet{kr72} proved the existence of the critical inclination in the framework of
the restricted circular three-body problem for both situations: outer perturber
and inner perturber, which are $i \sim 39^{\circ}$ and $i \sim 63^{\circ}$ respectively, for the case that the ratio of the semimajor axis of
the inner body with respect to the outer body tends to zero.

Just for completeness, the Kozai dynamics was also studied in extrasolar systems as a natural extension of the
original asteroidal problem, that means, the planet that generates the perturbation is outside the perturbed planet,
see for example \citet{in97},
\citet{wh97}, \citet{holm97},
\citet{mt98} and
\citet[chapter 9]{vk06}, while the mutual perturbations of two-planet systems were only considered more recently in the context of the three body problem
\citep{mich2006,libe2009,mg2010}.
It is worth mention that a big deal of the analytical studies of the Kozai dynamics applied to satellites, small bodies, and exoplanets were developed assuming an \textit{external} perturber, being the case with a perturber inside the perturbed orbit, as occurs in the TNR, only analyzed with detail more recently \citep{mg2010,fl2010,mg2011}.

In the present work, following the framework of the model given by \citet{tm96} and \citet{mt96}, we start extending in Section \ref{nmodel} their study from Neptune region to large orbital $a$, $q$ and $i$ finding the possible limits of perihelion variations assuming a secular non resonant evolution and, especially, studying in detail the Kozai dynamics around
$i \sim 62^{\circ}$.
The results in form of energy level curves are presented in terms of $(\omega,q)$ for an immediate comprehension of
the possible perihelion variations.
In Section \ref{amodel} we present a secular analytical model that describes approximately the secular dynamics in the far solar system given an explanation for the oscillations around $i \sim 62^{\circ}$ and predicting new orbital behaviors.
In Section \ref{reso} we analyze how a MMR enhances the KR strongly affecting the energy level curves and, especially, the peculiar 1:N resonances with Neptune.
In Section \ref{val}, analyzing a series of numerical integrations of fictitious particles with semimajor axes from 50 to 500 AU we discuss
the applicability of the secular Kozai dynamics and
the conditions for the capture in the MMR+KR
and for the enhancement of the perihelion distances.

\section{Non resonant Kozai dynamics beyond Neptune}
\label{nmodel}

We followed the method described in \citet{tm96} but including all the planets from Mercury to Neptune assuming coplanar and
circular orbits, as is usual. The inclusion of the terrestrial planets is not relevant for the study of the TNR but we have included them in order to study particular cases with $q$ inside the inner planets if necessary. In this context
we numerically calculated the double integral defined by the mean of the perturbing function over mean anomalies of the planets and the test particle and obtained
a mean perturbing function which is independent of the particle's longitude of the ascending node, $\Omega$, and mean anomaly, $M$. This implies that $a$  and
$H=\sqrt{1-e^2}\cos i$ are constant as explained in the reference above.
We constructed energy level curves  $K(\omega,q)$, where we preferred the perihelion distance instead of
$X=\sqrt{1-e^2}$ as in \citet{tm96} because it directly gives the path in $q$ that a TNO can perform.

\begin{figure}[]
\resizebox{13cm}{!}{\includegraphics{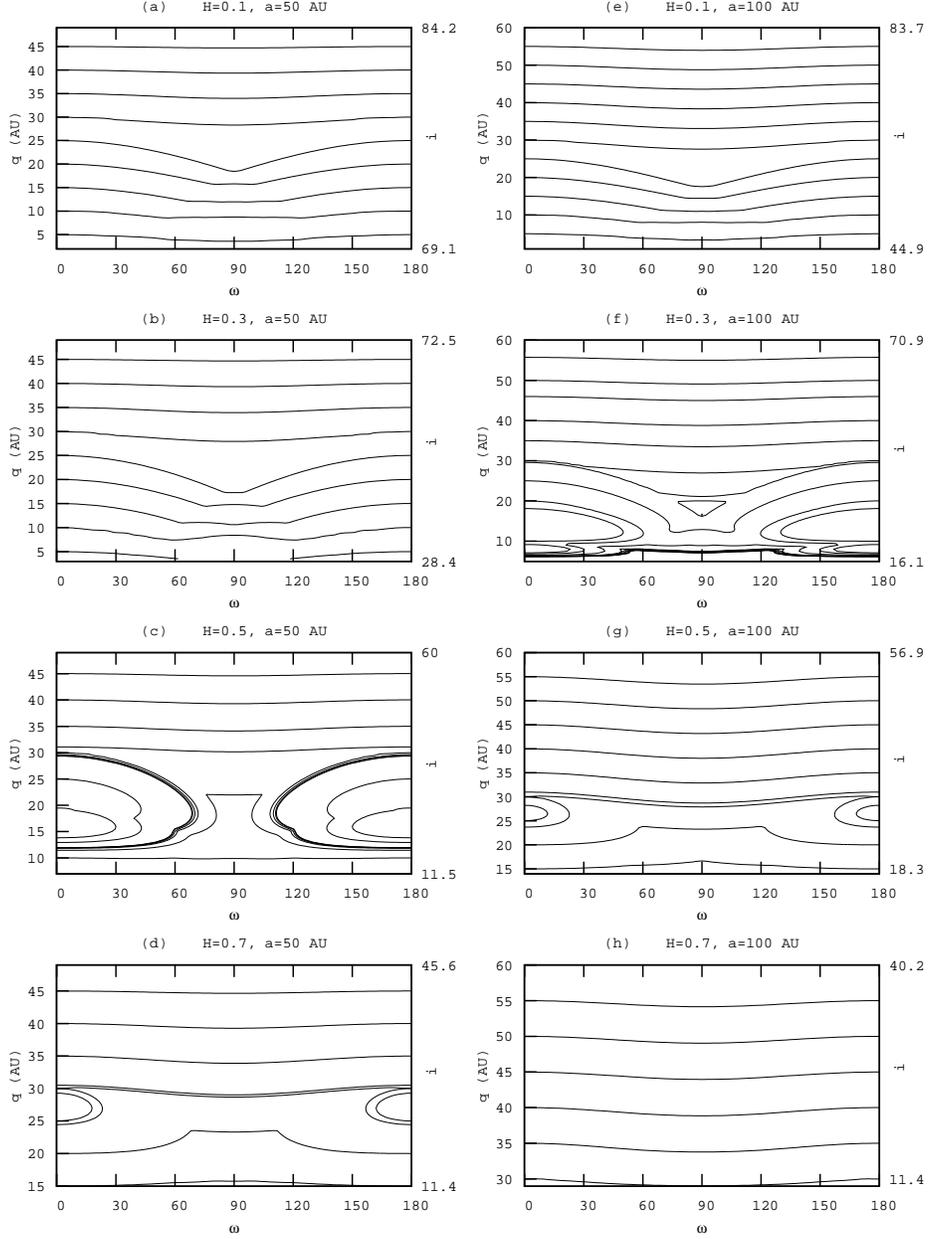}}
\caption{Energy level curves for $a=50$ and 100 AU for different values of $H$ showing the most relevant perihelion and orbital inclination variations
near Neptune and beyond. $q$ and $i$ are correlated by $H$ constant and the figures are symmetric with respect to $\omega = \mathrm{k}\, 90^{\circ}$. Note the equilibrium points located at $q<30$ AU and $\omega = 0^{\circ}, 180^{\circ}$. Inclination in degrees.}
\label{50100}
\end{figure}

\begin{figure}[]
\resizebox{13cm}{!}{\includegraphics{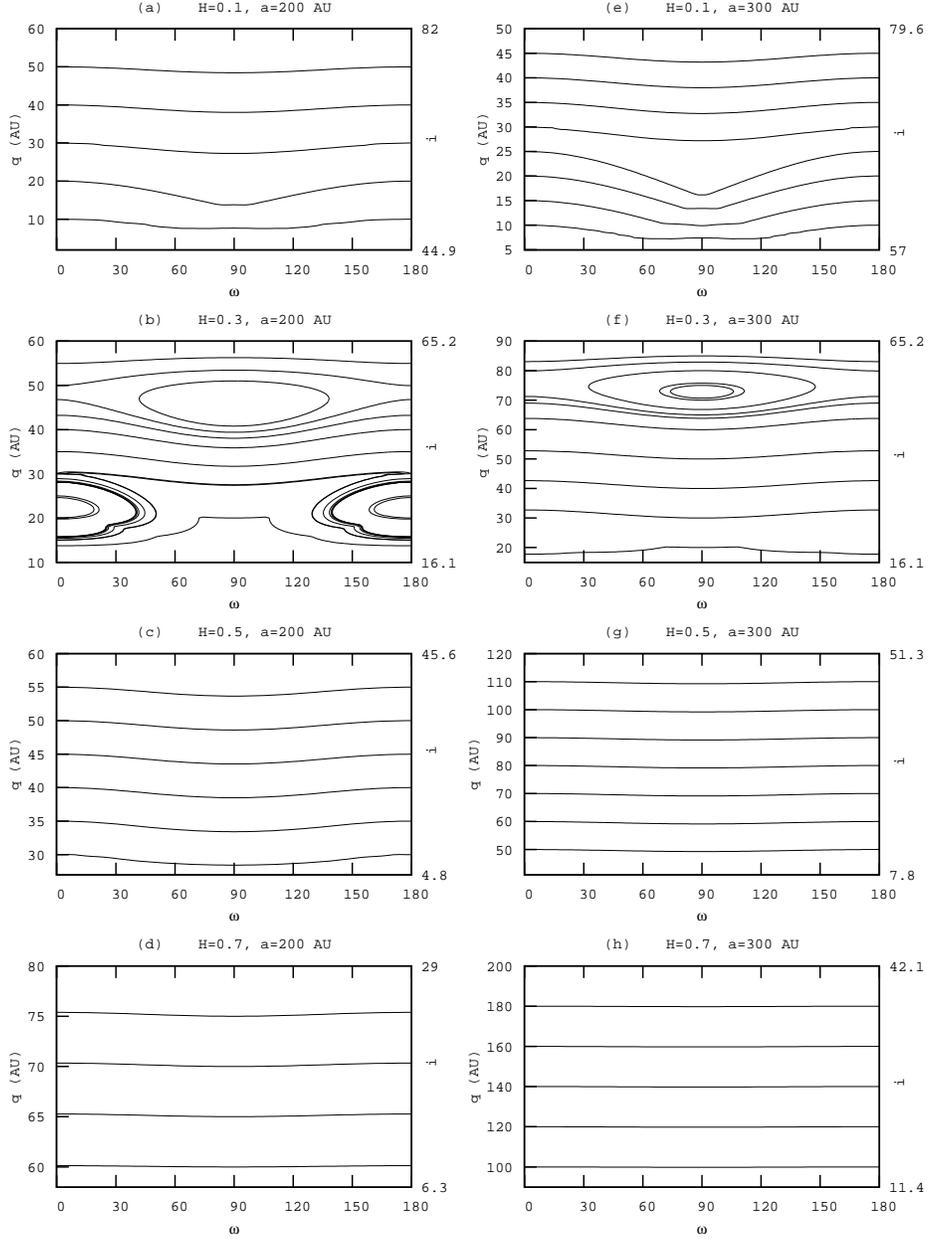}}
\caption{Same as Fig. \ref{50100} for $a=200$ and 300 AU. A new equilibrium point appears at $\omega = 90^{\circ}$ with a corresponding inclination of $\sim 63^{\circ}$.
For orbits with $H\geq 0.5$ no relevant orbital variations are allowed in this range of $q$.}
\label{200300}
\end{figure}

\begin{figure}[]
\resizebox{12cm}{!}{\includegraphics{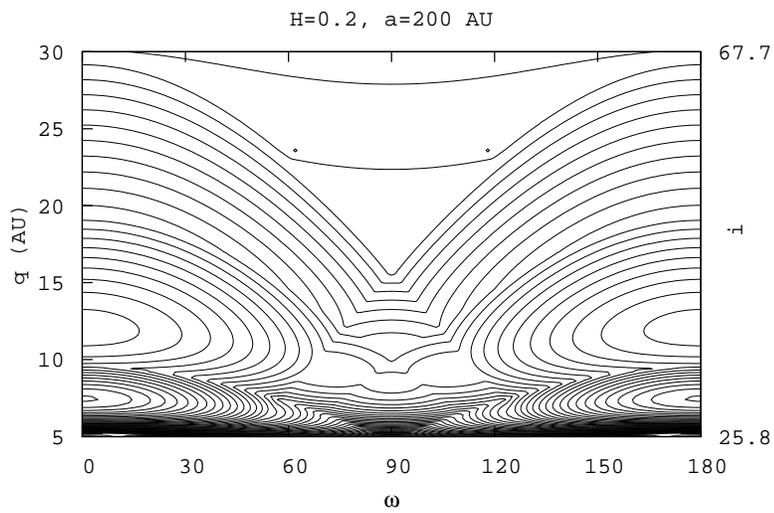}}
\caption{A typical energy level map in the region $q<30$ AU. Note the \textit{asymmetrical} equilibrium islands at $\omega \sim 70^{\circ}$ and $110^{\circ}$, between the collision curves with Neptune, Uranus and Saturn.}
\label{qlt30}
\end{figure}

In Figs. \ref{50100} and \ref{200300} we show a synthesis of energy level curves for $H=0.1, 0.3,
0.5$ and 0.7 and for $a=50, 100, 200$ and 300 AU. They are symmetric with respect to
$\omega = \mathrm{k}\, 90^{\circ}$, and $i=90^{\circ}$ so
we limited the plot to the region $0^{\circ}\leq \omega \leq 180^{\circ}$ and $i<90^{\circ}$, or $H>0$.
As $H \rightarrow 1$ (or $e, i \rightarrow 0$) the energy level curves are
progressively more  straight horizontal lines indicating that the allowed variations in $e,i$
become negligible.
The structure for $q<30$ AU showed by Figs.  \ref{50100} and \ref{200300} is well known since the work of \citet{tm96}.
Just for illustration, we show in Fig. \ref{qlt30} a typical map of energy level curves in the region interior to the giant planets.
Although the origin of some centaurs
could be explained with this dynamics, due to the energy level curves connecting $q\sim a_N$ with  $q < a_N$,
 we remark that, analyzing all ours energy level diagrams that cover a wide range of $q$ in the TNR, we conclude that, in general, there is no connection between $q>30$ AU and $q<30$ AU. That means,
this secular, non resonant, dynamics does not allow stable perihelion variations that cross Neptune's region. There exist some trajectories
that enter the region $q<30$ AU but inevitably reach the collision curve with Neptune
and in that case the method is not more valid, the energy is not more conserved and the motion will not
 follow the energy level curves.
 Nevertheless, we will see later in Section  \ref{reso} that the boundary  $q\sim 30$ AU can be crossed in a smooth regular orbital
 evolution following the energy level curves in
 the case of TNOs captured in MMRs with Neptune as showed for example by \citet{go05}.

\begin{figure}[]
\resizebox{13cm}{!}{\includegraphics{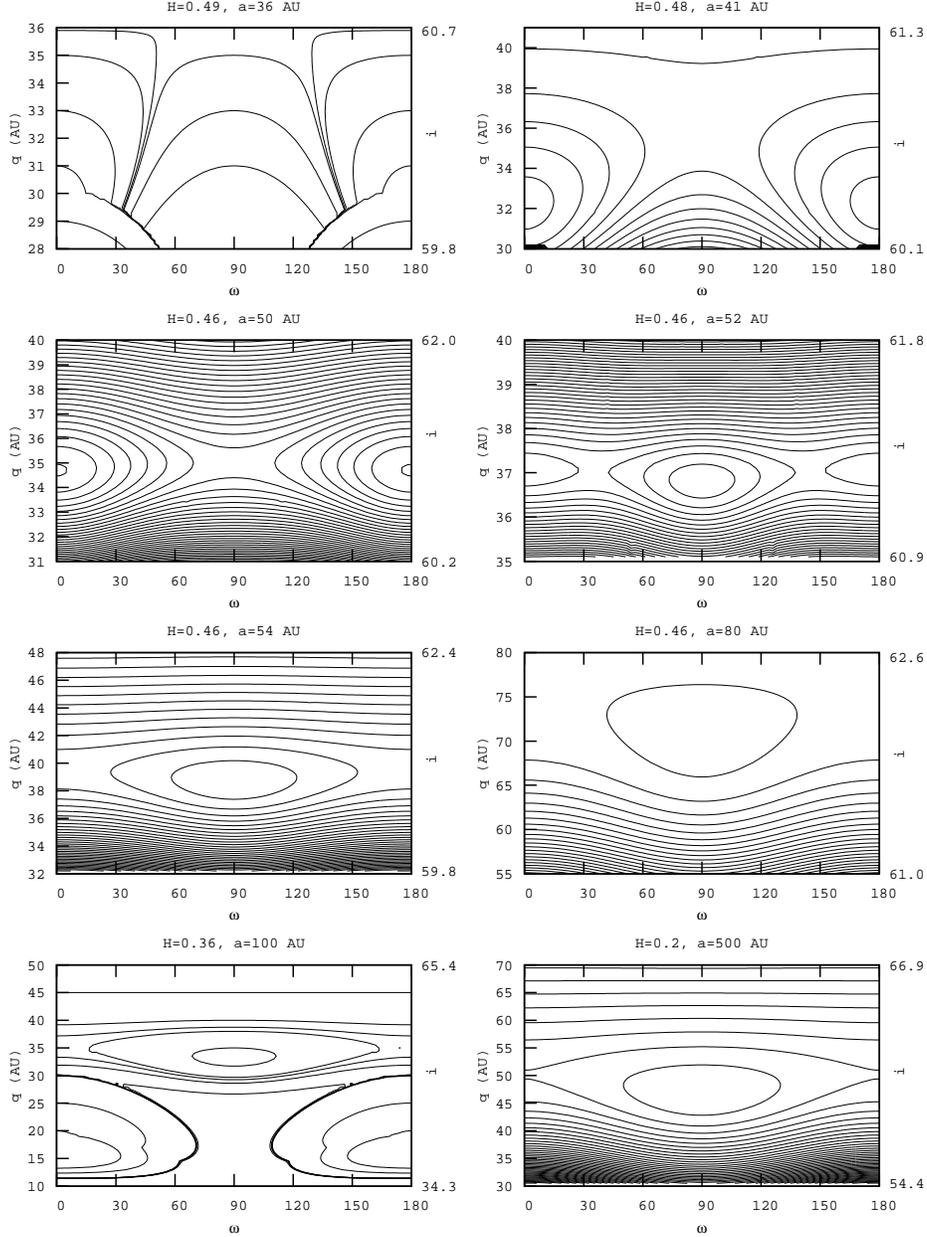}}
\caption{Metamorphosis of the Kozai resonance for $i \sim 61^{\circ} - 63^{\circ}$, the \textit{critical inclination}, outside Neptune's orbit. At $a=36$ AU all trajectories with $q>30$ AU and $H=0.49$ end collapsing in the collision curve with Neptune. From $a=41$ AU there appear
equilibrium islands at $\omega=0^{\circ}, 180^{\circ}$ similar to the ones obtained by \citet{ku02}. At $a=52$ AU a new equilibrium point appear at $\omega=90^{\circ}, 270^{\circ}$.
At $a=54$ AU the equilibrium points at  $\omega=0^{\circ}, 180^{\circ}$ disappear  and from then on only survive the equilibrium point at $\omega=90^{\circ}, 270^{\circ}$.
This Kozai resonance generates variations of $\sim 10$ AU in $q$ and it is confined to $i \sim 61^{\circ} - 63^{\circ}$.}
\label{metam}
\end{figure}

\begin{figure}[]
\resizebox{12cm}{!}{\includegraphics{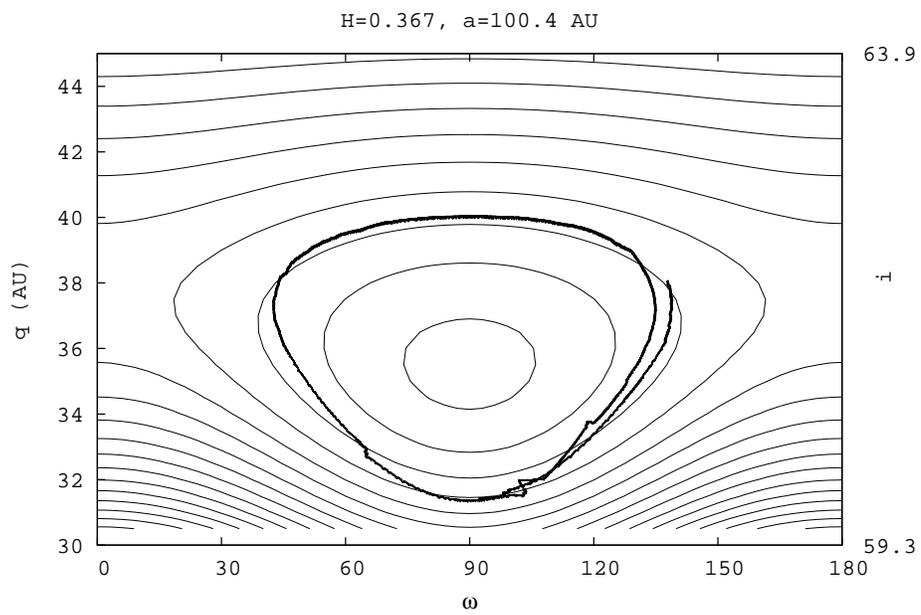}}
\caption{Example of secular long term variation in perihelion distance due to Kozai resonance
around the critical inclination ($i \sim 63^{\circ}$).
Points correspond to a numerical integration of a fictitious particle by 1 Gyr.}
\label{critnum}
\end{figure}

We also note that in the region $q>30$ AU in general there are not relevant variations in $q$ except for objects with $H<0.5$
where an impressive island of equilibrium appears at $\omega = 90^{\circ}$ and $270^{\circ}$, clearly shown in Fig. \ref{200300} in a narrow range of inclinations
around $63^{\circ}$. By fine tuning the parameters $(H,a)$ we reconstructed the transformations that these equilibrium islands experience from $a=36$ to $a=500$ AU which are showed in Fig.
 \ref{metam}. At the lower values of $a$ for which the equilibrium points exist  the corresponding inclination is $\sim 61^{\circ}$ and from $a>100$ AU the inclination of the equilibrium point is $\sim 63^{\circ}$.
The
perihelion variations that they generate can be of about 10 AU or more, and always occurs for $q$ outside Neptune.
Fig.  \ref{metam} shows that in fact the birth of the equilibrium points occurs for $a\sim 36$ AU and are located at $i\sim 61^{\circ}$, $\omega = 0^{\circ}, 180^{\circ}$.
 For larger $a$ they appear new equilibrium islands at  $i\sim 61^{\circ}$, $\omega = 90^{\circ}, 270^{\circ}$ which are the only ones that survive for $a\gtrsim 54$ AU.
Equilibrium points around $\omega=90^{\circ}$ are present
in previous works
but for very different, lower inclinations and very low perihelion values, well inside the planetary region, contrary to the case showed in our Figs. \ref{200300} and \ref{metam}.
In Fig. \ref{critnum} it is showed a fictitious particle evolving in the TNR obtained from a numerical integration of the outer Solar System
covering 1 Gyr using EVORB \citep{fgb02}
and the corresponding energy level curves in a good match.
Below the equilibrium points at $i \sim 63^{\circ}$ it is verified $d\omega/dt>0$ and above,   $d\omega/dt<0$.
These plots explain the previous results by \citet{ku02} and \citet{ga06}.
It is evident that the oscillation of $\omega$, which defines the Kozai resonance (KR), is installed in all the TNR at $i \sim 61^{\circ} - 63^{\circ}$, constituting a family
of equilibrium points in the Kozai dynamics.

From our maps we can conclude there is a connection between perihelion near Neptune and
very low perihelion, which is a well known result from previous studies, and,
more important,
there is another connection between perihelion near Neptune and perihelion outside Neptune's orbit through a new equilibrium point located at $i\sim 62^{\circ}$.
We remark there is no connection between both regions, that means, from the outer region to the inner one. Neptune's orbit is
like a barrier between these regions.
In fact, the chaotic evolution associated to orbits with $q<36$ AU \citep{du95,gl02,fgb04}  allows the
connection between both regions, but not more under the hypothesis of a secular regular evolution.
We will show in Section \ref{reso} that this scheme is strongly modified for particles captured in MMR with Neptune, but first, in the next section
we will find a dynamical justification for the equilibrium points at the critical inclination.

\section{An analytical secular non resonant approach}
\label{amodel}

We will obtain a simple secular model
which is in essence the same proposed by \citet{ko62} but applied to the TNR and taking into account
the perturbations due to $N$ planets moving in circular and coplanar orbits.
The secular evolution of a TNO with perihelion outside Neptune's orbit can be approximately modeled by the perturbative effects of a series of
massive belts representing the $N$ planets. For each planet $j$ we define a belt in the plane (x,y) with radius equal to the semimajor axis
of the planet, $a_j$, and with mass equal to the planet's mass $m_j$ in units of solar masses.
Following \citet[chapter III]{bc64} it is possible to show this model generates a perturbing function given by:

\begin{equation}
\label{r}
    R = \frac{C}{4}\frac{\mu}{r^{3}}(1-3\frac{z^2}{r^2}) + \frac{9E}{64}\frac{\mu}{r^{5}}(1-10\frac{z^2}{r^2}+\frac{35}{3}\frac{z^4}{r^4})  + \ldots
\end{equation}
where $C$ is the moment of inertia in the direction of $z$:
\begin{equation}
\label{c}
C=\sum_{j}^{N} m_j a_{j}^{2}
\end{equation}
and
\begin{equation}
\label{ce}
E=\sum_{j}^{N} m_j a_{j}^{4}
\end{equation}
being $\mu=k^2$, the square of the Gaussian constant, and $z$ and $r$ are z component of the TNO and its heliocentric distance respectively. Actually, the planets and also the Sun considered as a solid body contribute
to $C$ and $E$ but the contribution due to the Sun is negligible compared with the contribution of the planets, then we discard the Sun's contribution
to the perturbing function. Taking into account the four major planets we obtain $C=0.1161$ $M_{\odot} \mathrm{AU}^{2}$ and $E=52.0057$  $M_{\odot} \mathrm{AU}^{4}$.
The ratio between the term depending on $E$ respect to the term depending on $C$ is about $\sim 300/r^2$ which for a
typical TNO is less than 0.1, but we cannot neglect it.
Calculating the mean respect to mean anomaly of the perturbed TNO we get the mean perturbing function $R_m$:
\begin{equation}
\label{rm}
R_m=\frac{\mu}{16 a^3 \left( 1-e^2 \right)^{3/2}} (R_2+R_4) + \ldots
\end{equation}
where
\begin{equation}
\label{r2}
R_2 = C   (1+3 \cos 2i)
\end{equation}
and
\begin{equation}
\label{r4}
R_4=\frac{9  E   \left( \left( 2+3 e^2 \right) (9+20 \cos 2i +35 \cos4i)+40 e^2 (5+7 \cos2i) \cos 2\omega
\sin^2 i \right) } {512 a^2 \left( 1-e^2 \right) ^{2}}
\end{equation}
being $a,e,i$ the TNO's orbital elements.
For a typical TNO, $R_4/R_2 < 0.1$ and vanishes for growing values of $a$.
Using Laplace planetary equations \citep{md99} we obtain the secular evolution of the TNO:
\begin{equation}
\label{adot}
    \frac{da}{dt}=0
\end{equation}

\begin{equation}
\label{edot}
    \frac{de}{dt}=\frac{45 e k E}{512 a^{11/2}(1-e^{2})^3}(5+7\cos 2i) \sin^2 i \sin 2\omega + \ldots
\end{equation}

\begin{equation}
\label{idot}
    \frac{di}{dt}=-\frac{45 e^2 k E}{1024 a^{11/2}(1-e^{2})^4}(5+7\cos 2i) \sin 2i \sin 2\omega + \ldots
\end{equation}

\begin{equation}
\label{ndot}
    \frac{d\Omega}{dt}=-\frac{3 C k}{4 a^{7/2}(1-e^{2})^2}\cos i + \ldots
\end{equation}

\begin{equation}
\label{wdot}
    \frac{d\omega}{dt}=\frac{3 C k}{16 a^{7/2}(1-e^{2})^2}(3+5\cos 2i) + \ldots
\end{equation}
and from Eqs. (\ref{ndot}) and (\ref{wdot}):
\begin{equation}
\label{pdot}
    \frac{d\varpi}{dt}=\frac{3 C k}{16 a^{7/2}(1-e^{2})^2}(3 - 4\cos i + 5\cos 2i) + \ldots
\end{equation}
where only the lower order terms are showed.
Perturbations for $\Omega, \omega, \varpi$ are due to the lower order terms of the perturbing potential,
that means, the terms depending on $C$, while the perturbations for $i,e$ are due to higher order terms of the perturbing potential
depending on $E$.
Note that $di/de=e(1-e^2)^{-1}/\tan i$ which is the same relationship we can deduce from
$\sqrt{1-e^2}\cos i = H$, where $H$ is constant.

From these equations it follows that particles with orbits verifying $\cos(2i) \sim -3/5$  (or $i \sim 63^{\circ}$ and $117^{\circ}$) will verify $d\omega/dt \sim 0$ while, in general, $de/dt \neq 0$ and $di/dt \neq 0$.
In an energy level diagram $(\omega,q)$ this evolution corresponds with a trajectory which has a vertical tangent ($\omega =$ constant) when that inclination is reached, indicating a returning point
in the energy level curves. In particular, at
$\omega=\mathrm{k}\, 90^{\circ}$ according to the equations above we also  have $de/dt = di/dt = 0$, obtaining four equilibrium points, two stable and two unstable.
In analogy to the motion of an artificial satellite there is a \textit{critical inclination}, $i_c$,
 \citep[chapter 3]{be2} \citep[chapter 3]{mc05} which produces $d\omega/dt=0$. Above $i_c$ the argument of the perihelion
 has a retrograde motion and
below $i_c$ a prograde motion.
This is in good agreement with the result obtained by \citet{ga06}
and with the value obtained  by \citet{kr72} using a different approach in the framework of the restricted circular three-body problem.

\begin{figure}[]
\resizebox{12cm}{!}{\includegraphics{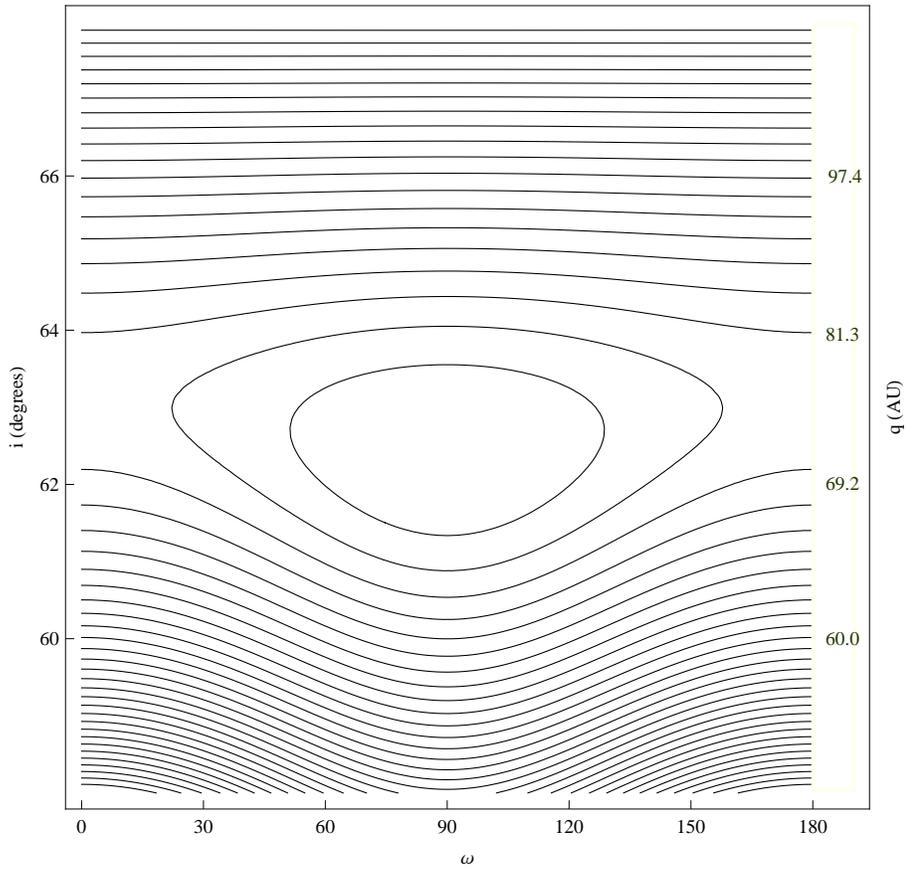}}
\caption{Energy level curves corresponding to a particle with $H=0.3$ and $a=300$ AU, the same as in
Fig. \ref{200300}(f) but in the space $(\omega,i)$ and calculated using the analytical model of Section \ref{amodel}.}
\label{critmodel}
\end{figure}

Substituting $e^2$ by its equivalent $1-(H/\cos i)^2$ in Eq. (\ref{rm})  we obtain $R_m$ depending on two variables
$(\omega,i)$ and two parameters $(a,H)$. Then, we can construct level curves of $R_m(\omega,i)$ which are in fact the energy level curves.
We construct level curves of $R_m(\omega,i)$ between two extreme values of the inclination:
\begin{equation}
\label{ilim}
\arccos \Bigg[ H \Bigg( \frac{q_{lim}}{a}(2-\frac{q_{lim}}{a})\Bigg) ^{-1/2} \Bigg] < i < \arccos [H]
\end{equation}
being the lower one the minimum inclination to avoid the region corresponding to perihelion lower than a limit value $q_{lim} \sim a_N$,
where the model cannot be applied and the higher one is due to the condition $e^2>0$. These energy level curves
are qualitatively equivalent to the $K(\omega,q)$ showed in Section \ref{nmodel}  and describe very approximately the secular dynamics in
the far outer solar system, say for $a>60$ AU. At Fig. \ref{critmodel}  we present an example of energy level curves calculated with this model, which
corresponds with one of the maps of Fig. \ref{200300}, showing the oscillations around the
critical inclination in a good match in spite of being two quite different procedures. For  $a<60$ AU the model reproduces qualitatively well the energy level diagrams showed in Fig. \ref{metam} with small differences in $i_c$ and $H$.

The values of $d\Omega/dt$ and $d\varpi/dt$ obtained from equations above are very small compared with the
fundamental frequencies of the solar system, then no secular resonances should appear for any eccentricity or
any inclination in the far TNR, which is an already know result since \citet{kz91}. But there is a particular situation of $d\varpi/dt \sim 0$ which occurs for $i\simeq 46^{\circ}$  and $i\simeq 107^{\circ}$, where between these values the longitude of the perihelion has retrograde motion and prograde otherwise. This is in good agreement with \citet{ga06} where it was found that $d\varpi/dt \sim 0$  at $i\sim 45^{\circ}$ for the particular case of TNOs with $a\sim 149$ AU.

\begin{figure}[]
\resizebox{12cm}{!}{\includegraphics{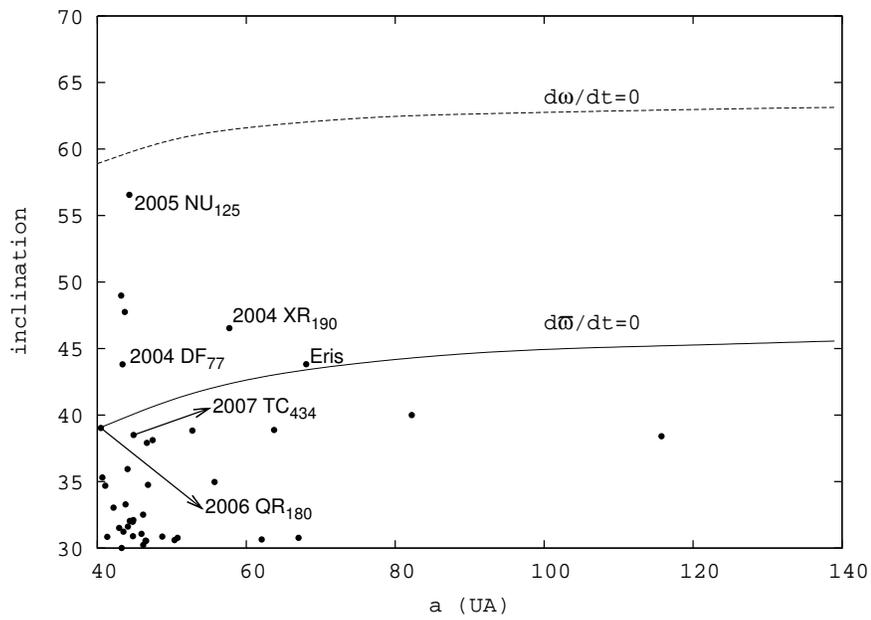}}
\caption{The critical inclination (dashed line) and the inclination which generates $d\varpi/dt=0$ (solid line)
as a function of the semimajor axis, according to the analytical model. The TNOs with $q>30$ AU are also plotted. }
\label{objects}
\end{figure}

\begin{figure}[]
\resizebox{12cm}{!}{\includegraphics{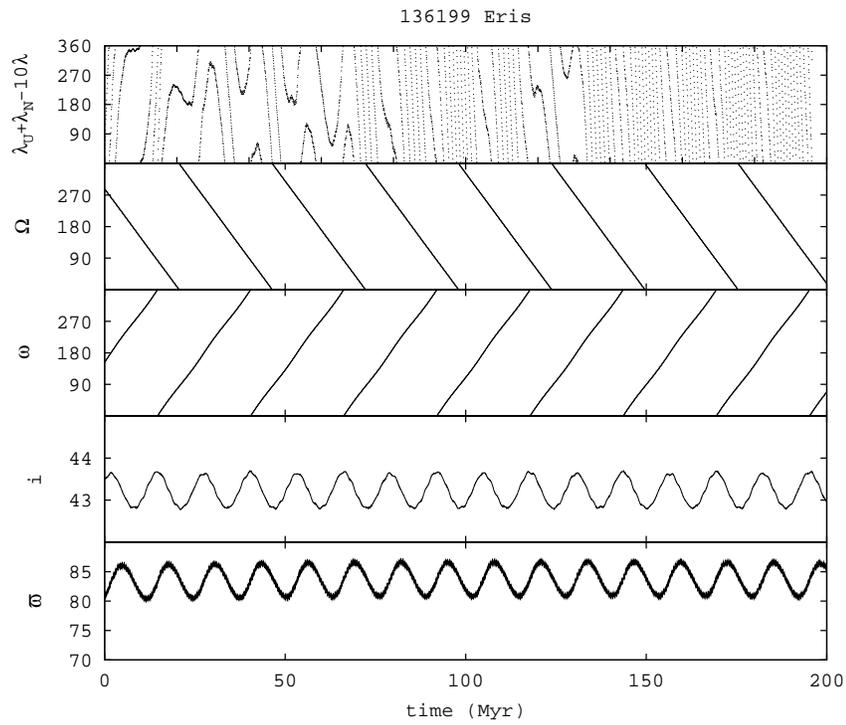}}
\caption{Orbital evolution of (136199) Eris. The orbital inclination of Eris is inside a range of orbital inclinations where the longitude of the perihelion
oscillates as predicted by our analytical model. Eris is also inside the three body resonance $\lambda_U + \lambda_N - 10\lambda \sim 0$. Orbital elements are referred to the invariable plane of the Solar System.}
\label{eris}
\end{figure}

More precisely, $i_c \sim 63^{\circ}$ and  $i \sim 46^{\circ}$ are values obtained from Eqs. (\ref{wdot}) and (\ref{pdot}) ignoring the higher order terms depending on $E$. If we take into account that terms these inclinations will show a dependence mainly with $a$, as showed in Fig. \ref{objects},
where we also plot the TNOs with $q>30$ AU and $i>30^{\circ}$ which are near these inclinations.
The object 2005 NU$_{125}$ is near to the critical inclination but it is not captured around  the equilibrium point, it has  $d\omega/dt > 0$.
The objects 2004 XR$_{190}$ and 2004 DF$_{77}$ are close to the curve that annulate $d\varpi/dt$ but not enough to have their perihelion
oscillating around a fixed value, on the contrary they have  $d\varpi/dt < 0$.
The object 2007 TC$_{434}$  has  $d\varpi/dt > 0$. Finally,
the objects 2006 QR$_{180}$ and Eris are almost exactly at the curve that annulate $d\varpi/dt$ according to our analytical model.
The first one exhibits large amplitude chaotic oscillations of its $\varpi$, but in the case of Eris, its $\varpi$
evolve oscillating quite regularly as is showed in Fig. \ref{eris}.
From the numerical integration of Eris we obtain a circulation period of its $\Omega$ and $\omega$ of  25.8 Myr while
our analytical model predicts 26.0 Myr. We note that Eris, the most massive known TNO, is the only one well captured in this
kind of motion, which can be considered as a resonance between the circulation frequency of its $\Omega$ and $\omega$.
An analogue resonance between the circulation frequency of its $\Omega$ and $\omega$ was found by \citet{libe2012} for an inner particle perturbed by an eccentric giant planet with mutual inclination of $\sim 35^{\circ}$, but in this case the resonant angle is $\omega - \Omega$,
not $\varpi = \omega + \Omega$ as in the case of Eris. However, the dynamics of Eris seems to be more complicated because it is very near to the
three body resonance $\lambda_U + \lambda_N - 10\lambda \sim 0$ involving Uranus and Neptune as showed in top panel of Fig. \ref{eris}.
Note  that this type of secular evolution should take place also in extrasolar planetary systems with an external inclined low mass planet
or even in a  binary system with an external inclined planetary companion.

\section{Kozai dynamics inside mean motion resonances with Neptune}
\label{reso}

When imposing a resonant condition with Neptune the energy level diagrams change completely.
We can understand the transformation the dynamics experiences when a MMR with Neptune is installed using a classical
disturbing function as in \citet{ga06}.
Considering the secular and resonant terms due to Neptune in the disturbing function of a particle in the TNR, the equation for the time evolution
of the eccentricity, for example, takes the form
\begin{equation}
\label{epsecres}
\frac{de}{dt}=f_2 \sin (2\omega)  + \ldots + f'_0 \sin (\sigma) + f'_1 \sin (\sigma)\cos (2\omega)
- f'_1 \cos (\sigma)\sin (2\omega)
+ \ldots
\end{equation}
where $\sigma$ is the principal critical angle of the resonance and the $f$ and $f'$ are coefficients depending on $a,e,i$. There is a similar expression for $di/dt$.
The critical angle is defined as
\begin{equation}
\label{sigma0}
 \sigma = (\mathrm{p+q})\lambda_N -\mathrm{p}\lambda -\mathrm{q}\varpi
 \end{equation}
where $\lambda_N, \lambda$ are the mean longitudes of Neptune and the particle and the integers $\mathrm{q,p}$ are the order and degree of the resonance respectively.
In the case  of non resonant motion,
 the angle  $\sigma$ is a fast circulating angle, then in mean it does not contribute to the time evolution of $e$ and $i$
 making all terms depending on the $f'$ to vanish
 and the only remaining contribution is due to $f_2 \sin (2\omega)$, the same we obtained with our secular model of Section \ref{amodel}, where $\omega$ is a slow circulating or oscillating angle.
But, in case of resonant motion, the angle  $\sigma$ should be considered as a fast oscillating angle, and its contribution to the
time evolution of $e$ and $i$ depends on the libration center, $\sigma_0$, and its amplitude. This makes an important point, as we explain below.

From the Eq. (\ref{epsecres}) for $de/dt$ and its analogue for $di/dt$ it is possible to deduce that there should be a different behavior between a pure secular evolution (terms with $f'$ disappear) and secular plus resonant evolution (terms with $f'$ show up). Moreover, there should be a difference between exterior MMRs of the type 1:N and other MMRs because in the resonances of the type 1:N the libration center of the critical angle is $\sigma_0 \neq 0^{\circ}, 180^{\circ}$,
and that means the terms depending on $\sin (\sigma)$ in mean do not vanish whereas in the case of the other resonances they do.
Summarizing, the  principal terms for each type of motion are the following.
For secular evolution:
\begin{equation}
\label{epsecular}
\frac{de}{dt}=f_2 \sin (2\omega)
\end{equation}
for MMRs with $\sigma_0 = 0^{\circ}, 180^{\circ}$
\begin{equation}
\label{epsimm}
\frac{de}{dt}=(f_2 - f'_1 <\cos(\sigma)>) \sin (2\omega)
\end{equation}
and for asymmetric librations ($\sigma_0 \neq 0^{\circ}, 180^{\circ}$, resonances 1:N):
\begin{equation}
\label{epasimm}
\frac{de}{dt}=( f_2 - f'_1 <\cos (\sigma)>  )\sin (2\omega)  +  f'_0 <\sin (\sigma)> + f'_1 <\sin (\sigma)>\cos (2\omega)
\end{equation}
where $<\cos (\sigma)>$ and $<\sin (\sigma)>$ refer to the mean values over a libration cycle, which vanish for large amplitude librations converting both resonant cases in the secular one.
The two first equations give equilibrium points at $\omega = \mathrm{k}\, 90^{\circ}$ as is usual in secular motion.
But from Eq. (\ref{epasimm}) we conclude that the equilibrium points for resonances 1:N should be shifted from $\omega = \mathrm{k}\, 90^{\circ}$.

\begin{figure}[]
\resizebox{12cm}{!}{\includegraphics{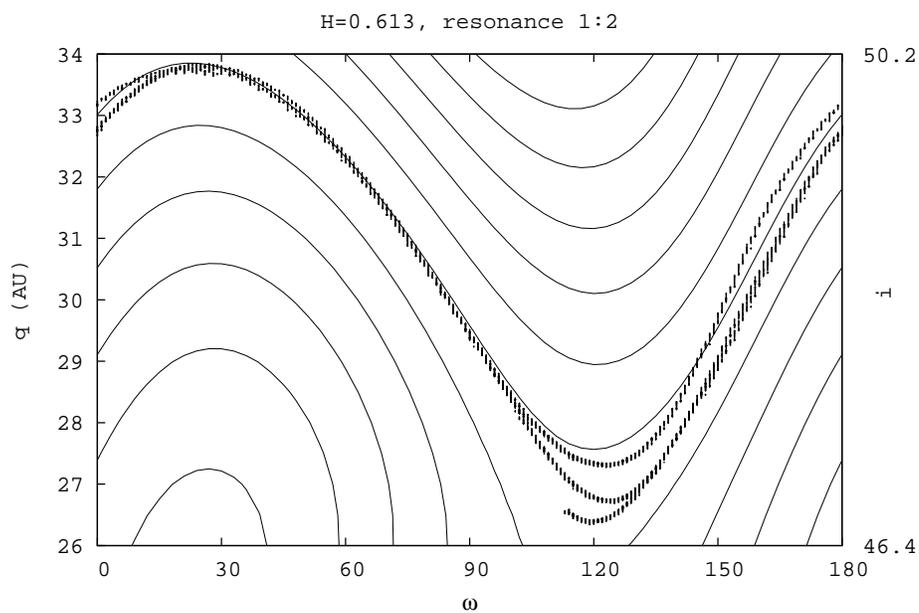}}
\caption{Points indicate the evolution in $(\omega,q)$ of a fictitious particle captured in resonance 1:2 with Neptune with libration center at $\sigma_0 \sim 295^{\circ}$ and
libration amplitude $20^{\circ}$ obtained by numerical integration in an interval of 10 Myrs. In the background are the energy level curves for a particle with the same $H=0.613$ and same $\sigma_0$ and amplitude as the particle.}
\label{res1to2}
\end{figure}

Energy level curves for a particle under the perturbation of the giant planets and assuming a resonant motion with Neptune can be
obtained
by the
method of Section \ref{nmodel}
but imposing
the resonant condition between the test particle and Neptune as in \citet{go05} and  \citet{go11}.
In the numerical calculation of the energy, $K$, we impose a link between $\lambda_N$ and $\lambda$ trough Eq. (\ref{sigma0}) according to an assumed time evolution of the resonant critical angle, $\sigma(t)$, which can be deduced from the resonant perturbing function as in \citet[see also http://www.fisica.edu.uy/$\sim$gallardo/atlas]{ga07}  or by numerical integrations.
We can calculate new
energy level curves imposing the resonant condition obtaining the evolution in the
space $(\omega,i)$ or $(\omega,q)$ and, as we have explained above, they will be very different from the secular non-resonant ones analyzed in Section \ref{nmodel}, allowing very different variation ranges for $q$ and $i$.
There are some examples in the literature \citep{go05,go08,go11} of energy level curves of fictitious particles captured in symmetric
resonances showing the equilibrium points at $\omega = \mathrm{k}\, 90^{\circ}$. We also have obtained several ones for different resonances and libration
amplitudes that confirm that behavior.
 To check the predicted behavior of the asymmetric resonances 1:N we constructed  several energy level curves for different MMRs.
For example, Fig. \ref{res1to2} shows the
evolution in the plane $(\omega, q)$ of a high inclination particle captured in 1:2 resonance with Neptune, librating with amplitude $\sim 20^{\circ}$ around a libration center located at $\sigma_0 \sim 295^{\circ}$, obtained from a numerical integration including all
the planets. In the background are the
 energy level curves corresponding to a particle with the same $H=0.613$ and imposing the same resonant condition
  that the particle shows.
  The numerical resonant model reproduces very well the exact numerical integration, with small departures mainly due to variations
  in the libration amplitude that the model does not take into account.
Note how the symmetry respect to $\omega = \mathrm{k}\, 90^{\circ}$ that the secular energy level maps exhibit is clearly broken.

\begin{figure}[]
\resizebox{13cm}{!}{\includegraphics{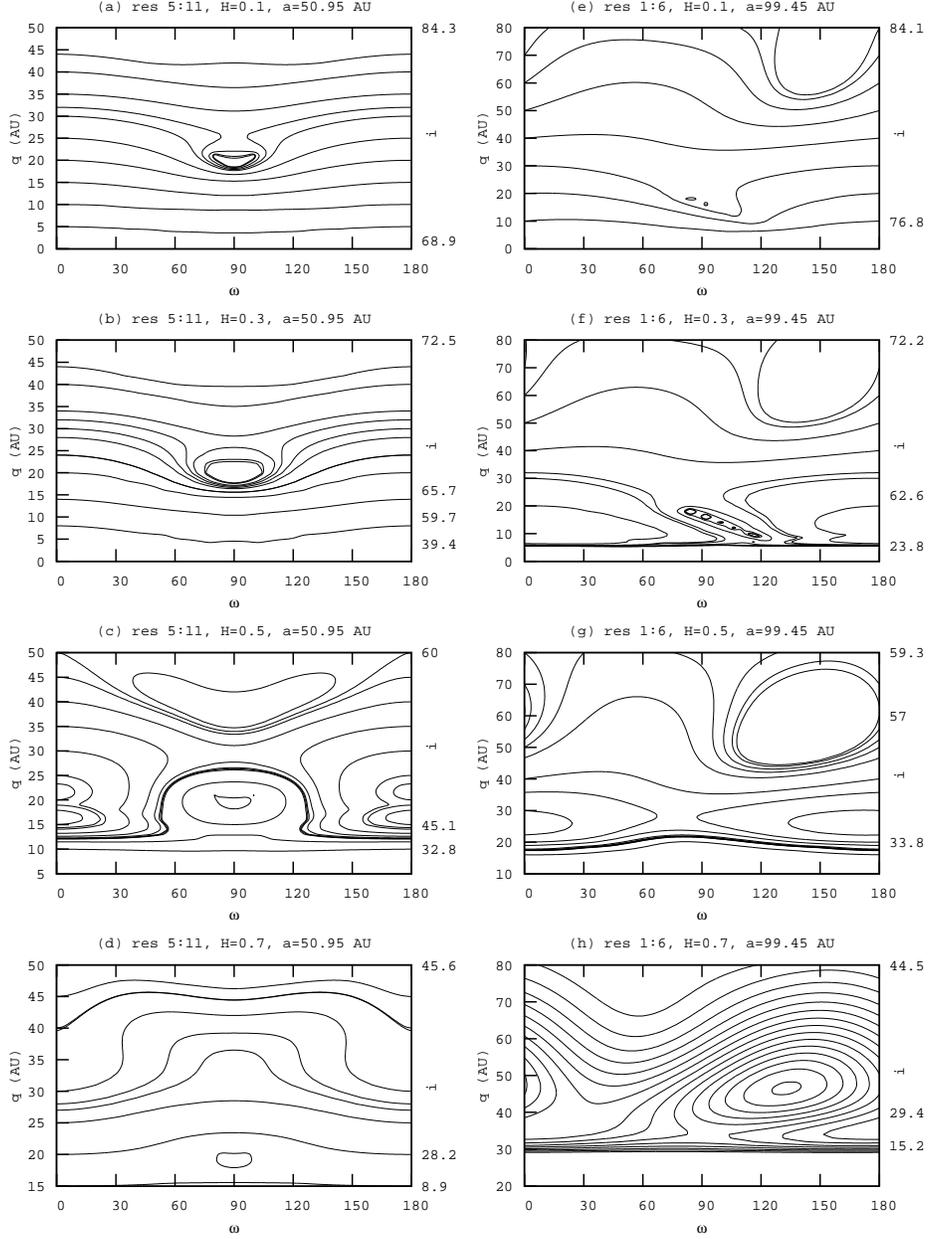}}
\caption{Energy level curves for orbital configurations similar to Fig. \ref{50100} but imposing a resonant condition. From (a) to (d) correspond to resonance 5:11 assuming $\sigma_0=180^{\circ}$
and amplitude $20^{\circ}$. From (e) to (h) correspond to resonance 1:6 assuming $\sigma_0=60^{\circ}$
and amplitude $20^{\circ}$. Compare with Fig. \ref{50100}. For growing libration amplitude the energy level curves tend to the secular non resonant case.}
\label{8resonances}
\end{figure}

Figure \ref{8resonances} corresponds to almost identical conditions as in Fig. \ref{50100} but imposing a resonant condition corresponding to resonance 5:11 (a-d) located at $a=50.95$ AU and
1:6 (e-h) at $a=99.45$ AU. Several features can be noted in this figure. First, the resonant energy level maps change drastically with respect to
the non resonant case, especially allowing large perihelion variations not showed in the non resonant case. Second, the symmetry
respect to $\omega = \mathrm{k}\, 90^{\circ}$ is broken in the case of resonances 1:N. This asymmetric behavior of the energy level curves is similar to the one obtained by first time by \citet{ko85} in his study of Jupiter's trojans.
Third, there are connections between perihelion
outside Neptune with perihelion inside Neptune's orbit, breaking Neptune's barrier of the non resonant case. Fourth, the islands corresponding to
the critical inclination are much wider and shifted from $63^{\circ}$.

\begin{figure}[]
\resizebox{12cm}{!}{\includegraphics{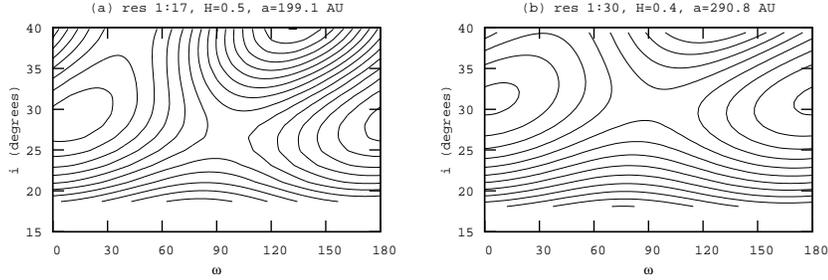}}
\caption{Energy level curves corresponding to resonance 1:17 and 1:30 assuming $\sigma_0=65^{\circ}$
and amplitude $20^{\circ}$ in both cases.
The minimum inclination necessary to connect low perihelion with high perihelion configurations
is greater for larger $a$.
}
\label{2resonances}
\end{figure}

It has been shown by means of numerical integrations \citep{gl02,go05,ga06,go08} and by the construction of some energy level curves
\citep{go05,go08,go11}
that the MMR+KR allows
high amplitude variations of the perihelion distance of SDOs only for inclinations above Pluto-like inclinations.
In the present work we can conclude that almost all substantial perihelion variations we observed in our energy level diagrams corresponding to
a MMR+KR dynamics occur when the orbital inclinations are above the minimum $i_{m}\sim 15^{\circ}$. Moreover, this minimum inclination necessary to connect low perihelion regions with high perihelion regions is greater for larger $a$ and it occurs for lower values of $H$. For example, in Fig. \ref{8resonances}(h) it is shown that for resonance 1:6 located at $a=99.45$ AU, $i_{m}$ is approximately $15^{\circ}$ with $H=0.7$ and in Fig. \ref{2resonances} the values that can be deduced from the plots are $i_{m}\sim 20^{\circ}$
for resonance 1:17 at $a=199.1$ AU  with $H=0.5$ and $i_{m}\sim 24^{\circ}$
for resonance 1:30 at $a=290.8$ AU  with $H=0.4$.

Summarizing, when imposing the resonance condition the energy level curves can cross the region $q \sim a_N$ connecting by constant energy level curves
the region $q > a_N$ and  $q < a_N$  allowing the biggest perihelion variations without suffering a close encounter with Neptune (Figs. \ref{res1to2} and \ref{8resonances}).
In particular, the MMR+KR dynamics can explain the origin of some objects with $q<a_N$ starting from $q>a_N$ without invoking close
encounters with Neptune. A pure secular, non resonant, evolution cannot generate such orbital transformations.
Resonances 1:N which are the stronger ones in the TNR exhibit asymmetric energy level maps.

\section{Kozai dynamics under diffusion in semimajor axis}
\label{val}

It has been shown that  TNOs with perihelion
distances $q\lesssim 36$ AU experience a diffusive process or a frankly random walk  in semimajor axis due to the
perturbations by the giant planets or by the mechanism of resonance
sticking contributing to the Scattered Disk Object population \citep{dl97,gl02} or even to the Oort cloud \citep{fgb04}.
That upper limit for $q$ is pushed to larger values for objects with larger $a$ \citep{gl02}.
  We re-analyzed this diffusive process
in a  wide range of $q,a$ and $i$ by means of numerical integrations of fictitious particles under the gravitational effect of the giant planets.
We took 1000 particles uniformly distributed in the phase space defined by
$0^{\circ}<i<70^{\circ}$,
$0^{\circ}<\Omega<360^{\circ}$,
$0^{\circ}<\omega<360^{\circ}$,
$50<a<500$ AU and
$31<q<f(a)$ where $f(a)= 40$ AU if $a<150$ AU and $f(a)=  30.0 + 0.085(a-30.0)$ AU if $a>150$ AU which is the region susceptible of  diffusion according to \citet{gl02}
and we followed them by 1 Gyr.

\begin{figure}[]
\resizebox{12cm}{!}{\includegraphics{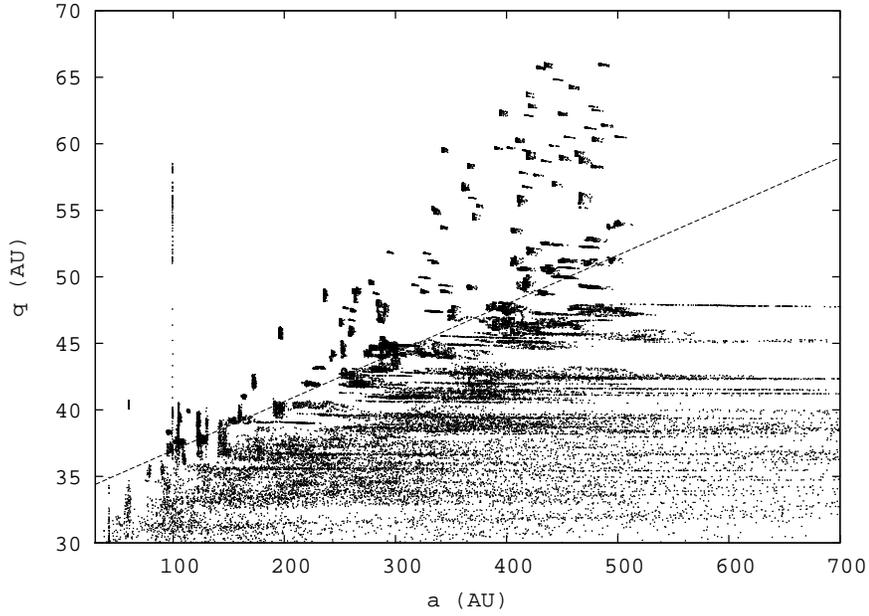}}
\caption{Superposition of the orbital states of a numerical integration by 1 Gyr of the outer Solar System plus 273 fictitious particles with initial $0^{\circ}<i<20^{\circ}$.
 The dashed line corresponds to $q = a/27.3 + 33.3$ AU which approximately defines the limit for the diffusion region in semimajor axis.
Diffusion in $a$ can be halted by a MMR but
once a capture in MMR occurs the Kozai dynamics does not generate relevant changes in $q$  due to the very low orbital inclination (with an extraordinary exception of a particle captured in resonance 1:6
at $a\sim 100$ AU).}
\label{difusion1}
\end{figure}

\begin{figure}[]
\resizebox{12cm}{!}{\includegraphics{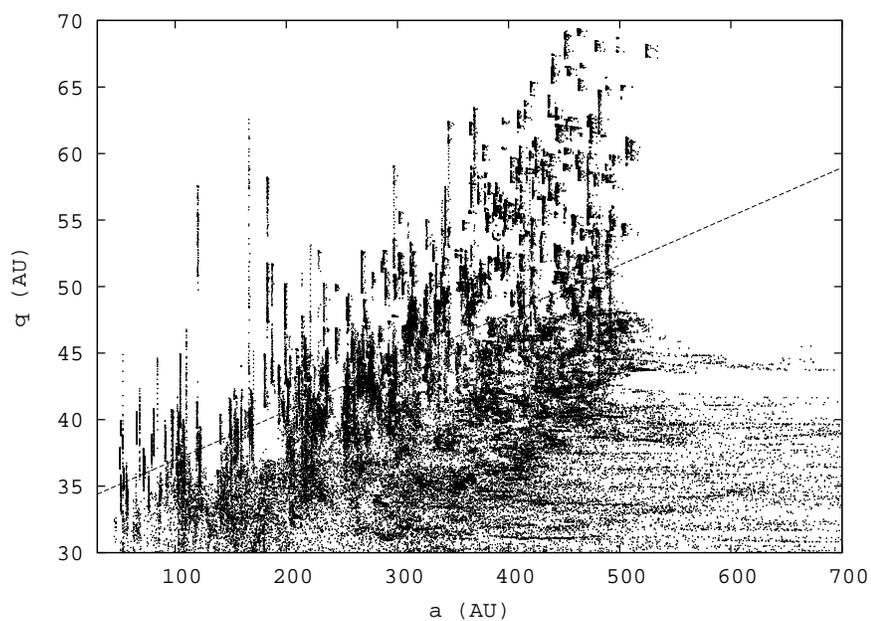}}
\caption{Superposition of the orbital states of a numerical integration by 1 Gyr of the outer Solar System plus 727 fictitious particles with initial $20^{\circ}<i<70^{\circ}$.
The diffusion process is frequently stopped by a capture in MMR+KR allowing large perihelion variations.
However, larger $a$ require larger $i$ to generate relevant perihelion variations  (see Fig. \ref{iakozai}). The dashed line is the same limiting line as in Fig.
\ref{difusion1}.}
\label{difusion2}
\end{figure}

We found the diffusion process in $a$ is mostly restricted to the region
$q \lesssim a/27.3 + 33.3$ AU
and in a lesser extent it is also depending on $i$:
small inclination orbits diffuse in a somehow wider region of $q$  than large inclination orbits.
But we found the orbital inclination is determinant in defining the time evolution of $q$:
almost all particles with initial $i<20^{\circ}$ in the region we studied  do not experience relevant perihelion variations.
In Fig. \ref{difusion1} we show in the plane $(a,q)$ the superposition of the orbital states of all particles with
initial $i<20^{\circ}$. Except for the very particular case  with  $a \sim 100$ AU there are no perihelion variations.
On the contrary,
the population with initial  $i>20^{\circ}$ shows evident variations in $q$ and this is illustrated in the
Fig.  \ref{difusion2} which was generated in same way as Fig.  \ref{difusion1}.
For low $i$, paths are almost horizontal with no variations in $q$. Meanwhile, for higher $i$
the diffusion in $a$ is in general alternated with  a migration in $q$ keeping $a$ constant, which indicates a capture in MMR+KR.

We analyzed  the captures in all resonances of the type 1:N up to 1:30 and 2:N up to 2:61 ($a \sim 294$ AU) and we conclude
that the capture in MMR is almost independent of the orbital inclination.
In our numerical integrations we detected approximately 210 captures in MMR 1:N and 2:N, with approximately 62\% of the captures corresponding to
1:N resonances and  38\% corresponding to 2:N resonances which is a ratio in agreement with the fact that resonances 1:N are stronger than resonances
2:N \citep{ga07,ly07}.
Once a capture in MMR occurs it stops the diffusion in semimajor axis and a secular Kozai dynamics is installed inside the MMR.
That secular dynamics does not allow relevant variations in the perihelion distances  except if the inclination is high enough as we have explained in Section \ref{reso}.

\begin{figure}[]
\resizebox{12cm}{!}{\includegraphics{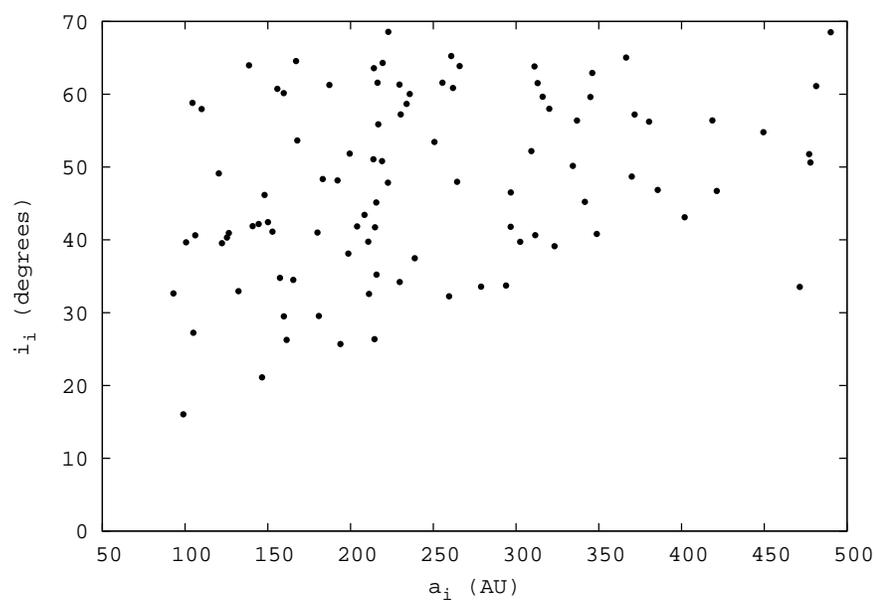}}
\caption{Initial $(a,i)$ for particles experiencing  perihelion variations  $\Delta q > 5$ AU   due to Kozai resonance. Orbits with larger $a$ require larger $i$ to generate relevant perihelion variations by means of MMR+KR.
Particles having close encounters with the planets were not considered in this plot, the inclination was taken with respect to the invariable plane of the Solar System and the semimajor axis is baricentric.}
\label{iakozai}
\end{figure}

 In Fig. \ref{iakozai} we show initial $(a,i)$
of the particles that experienced $\Delta q > 5$ AU due to MMR+KR, being discarded the cases of close encounters with the planets
and also the particles ejected from the solar system.
For larger $a$ the required inclination necessary
to generate relevant perihelion variations by means of MMR+KR
is larger. The values of the minimum inclination that we deduced from the energy level maps in Section \ref{reso}
are in agreement with the numerical results given by Fig. \ref{iakozai}.

\begin{figure}[]
\resizebox{12cm}{!}{\includegraphics{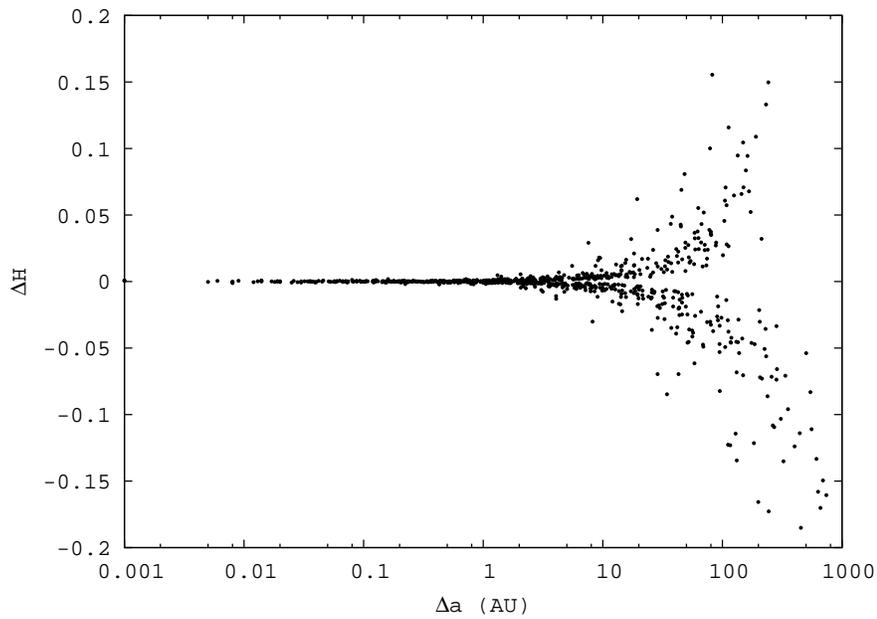}}
\caption{$H_{final} - H_{initial}$  versus $|a_{final} - a_{initial}|$ for particles surviving all the integration.
Particles with diffusion in $a$ up to few UAs conserve their $H$ but otherwise considerable variations
 in $H$ occur and
 the evolutive paths of these particles  will not be confined to the trajectories of $H=constant$.
 For the calculation of $H$ the inclination was considered with respect to the invariable plane of the Solar System.}
\label{deltaH}
\end{figure}

In the region under diffusion
regime in principle neither $a$ nor $H$ are conserved and consequently the methods of Sections \ref{nmodel} and \ref{reso} nor the analytical approach of Section \ref{amodel}
can be applied.
Analyzing the variations $\Delta H$ in our numerical integrations  we found that its conservation depends mostly on the
variations $\Delta a$ suffered by the particles (Fig. \ref{deltaH}).
But, the diffusion process can be stopped by the temporary capture in a MMR, in particular
the ones with enough strength and width, as resonances of the type 1:N or 2:N are \citep{ga06,ga07,ly07,go08}. Under the MMR regime the semimajor axis remains constant and the  models can be applied. In other words, in the diffusion region the models can be applied during the time intervals where the motion is dominated by MMRs.

Finally, we remark that our numerical integrations confirm
that $d\omega/dt \sim 0$ at $i_c \sim 63^{\circ}$ and
 $d\varpi/dt \sim 0$ at $i \sim 46^{\circ}$, as predicted by the analytical model.

\section{Conclusions}
\label{conc}

We  provide a series of energy level maps and analytical results describing the possible evolutive orbital paths for resonant and non resonant TNOs
driven by the Kozai dynamics from Neptune and beyond.
In the case of non resonant orbits beyond Neptune the KD does not generate relevant orbital variations except for orbits with
inclination near the critical value $i_c \sim 63^{\circ}$
 where variations of the order of 10 AU in $q$ are allowed and
$\omega$ oscillates around
$90^{\circ}$ or $270^{\circ}$ for $a>52$  AU and around
$0^{\circ}$ or $180^{\circ}$ for $a<52$ AU. These secular orbital variations always preserve the perihelion outside Neptune's orbit.
 We showed this secular KR is analogue to the
mechanism that makes $\omega$ to oscillate in a satellite orbiting an oblate planet.
Our analytical secular model also predicts that
TNOs with $i\lesssim 46^{\circ}$
or $i\gtrsim 107^{\circ}$ have a prograde motion of their $\varpi$ while orbits with $46^{\circ} \lesssim i \lesssim 107^{\circ}$ have retrograde evolution of their $\varpi$.
TNOs with orbital inclination near that limiting value should oscillate their longitude of the perihelion, which is the case of Eris.
The critical inclination and the inclination that annulate the time evolution of $\varpi$ have a dependence with $a$, especially for $a\lesssim 60$ AU.

In the case of resonant orbits, MMR+KR can generate large perihelion variations
if their orbital inclination is greater than a minimum value $i_m$ that depends on $a$: larger $a$ require larger $i_m$.
Contrary to the non resonant case, these orbital variations allow the perihelion to cross Neptune's orbit connecting the exterior planetary region to the interior
by a secular regular non-encountering dynamics.

The equilibrium points for the regime of MMR+KR are located at $\omega = \mathrm{k}\, 90^{\circ}$ and the energy level curves in the plane $(\omega,q)$
or $(\omega,i)$ are
symmetric with respect to these values of $\omega$ as in the non resonant case, except for the case of resonances of the type 1:N where
the symmetry is broken and the equilibrium points are shifted from the canonical values. This lost of symmetry is due to the fact that for resonances 1:N the libration center is not located at $\sigma_0 = 0^{\circ}, 180^{\circ}$ as in all others MMRs.

Particles with  $q \lesssim a/27.3 + 33.3$ AU are inside a diffusion region in $a$ where eventually could be halted by a capture in MMR.
If the diffusion in $a$ is greater than a few AUs, then $H$ can not be held constant and the particle will not follow trajectories
of equal energy. But, once the object is captured in a MMR the KD is installed and the evolutive paths follow curves of constant energy.

Starting with a primordial population with low inclination orbits it is not possible to reach
the present distribution of orbital inclinations of the TN population by means of the Kozai dynamics itself, being resonant or non resonant.
At least, a previous excitement mechanism is necessary to enhance the primordial inclinations allowing the Kozai dynamics to generate
a more diverse distribution of orbital inclinations and eccentricities in the TNR.

\bigskip

\textbf{Acknowledgments.}
 This study was developed in the framework of the projects
 ``Din\'{a}mica Secular de Sistemas Planetarios y Cuerpos Menores" (CSIC) and
``Caracterizaci\'{o}n de las Poblaciones de Cuerpos Menores del Sistema Solar" (ANII FCE 2007 318).
We acknowledge partial support by PEDECIBA. The authors acknowledge the suggestions given by the reviewers that with their criticism
improved the final version of this work.

\newpage

\end{document}